\documentclass[10pt,english,twocolumn,aps,prd,nofootinbib,preprintnumbers]{revtex4}
\usepackage{amssymb,amsmath,amsfonts,amsbsy,epsfig,graphicx}
\usepackage{multirow}
\usepackage{lipsum}
\usepackage{blindtext}
\usepackage{xspace}
\usepackage{pdflscape}
\usepackage{afterpage}
\usepackage{gensymb}
\usepackage{booktabs}
\usepackage[dvipsnames]{xcolor}

\usepackage{graphicx}
\usepackage{xcolor}
\usepackage{adjustbox}

\usepackage{amsmath}    
\usepackage{hyperref}

\usepackage{todonotes}

\newcommand{\be}{\begin{equation}}
\newcommand{\ee}{\end{equation}}
\newcommand{\bea}{\begin{eqnarray}}
\newcommand{\eea}{\end{eqnarray}}

\newcommand{\MM}{\mathcal{M}}
\newcommand{\DD}{\mathcal{D}}

\def\lcdm{$\Lambda$CDM }

\begin{document}

\title{Is there any measurable redshift dependence on the SN Ia absolute magnitude?}

\author{Domenico Sapone}
\email{domenico.sapone@uchile.cl}
\affiliation{Grupo de Cosmolog\'ia y Astrof\'isica Te\'orica, Departamento de F\'{i}sica, FCFM, \mbox{Universidad de Chile}, Blanco Encalada 2008, Santiago, Chile}

\author{Savvas Nesseris}
\email{savvas.nesseris@csic.es}
\affiliation{Instituto de F\'isica Te\'orica UAM-CSIC, Universidad Auton\'oma de Madrid,
Cantoblanco, 28049 Madrid, Spain}

\author{Carlos A. P. Bengaly}
\email{carlosap87@gmail.com}
\affiliation{D\'epartement de Physique Th\'eorique, \mbox{Universit\'e de Gen\`eve}, 1211 Gen\'eve 4, Switzerland}
\affiliation{Grupo de Cosmolog\'ia y Astrof\'isica Te\'orica, Departamento de F\'{i}sica, FCFM, \mbox{Universidad de Chile}, Blanco Encalada 2008, Santiago, Chile}

%\date{\today}
\begin{abstract}
\noindent We test the cosmological implications of a varying absolute magnitude of Type Ia supernovae using the Pantheon compilation, by reconstructing different phenomenological approaches that could justify a varying absolute magnitude, but also approaches based on cosmic voids, modified gravity models and a binning scheme. In all the cases considered in this work, we find good agreement with the expected values of the standard $\Lambda$CDM model and no evidence of new physics. 

\end{abstract}

\maketitle

\section{Introduction \label{sec:intro}}
One of the main problems in cosmology is to understand the apparent late-time acceleration of the Universe, first evidenced by the supernova surveys at the end of the previous century~\cite{Riess:1998cb, Perlmutter:1998np} and then by other experiments~\cite{Moresco:2015cya, Alam:2016hwk, Aghanim:2018eyx}. 
Within the standard cosmological principle, i.e. that the Universe is nearly isotropic and homogeneous, a component with a negative pressure $p<-\rho/3$ is required to drive the accelerated expansion, usually dubbed dark energy. However, several key questions about the nature of dark energy and why it should start to dominate the overall energy
density just now, currently remain unanswered \cite{Sapone:2010iz}. A completely different approach postulates that gravity needs to be modified at sufficiently large scales such that the late time acceleration is only apparent~\cite{Ferreira:2019xrr,Slosar:2019flp,DeFelice:2012vd,Nesseris:2010pc,Pantazis:2016nky} or that dark matter and dark energy may exhibit common features \cite{Kunz:2016yqy,Kunz:2015oqa}. Finally, a third approach even rejects the assumption of homogeneity and isotropy of the Universe and the dimming in the luminosity of the supernovae is attributed to the non-homogeneous structures in the Universe~\cite{Clarkson:2012bg}.

Current limits on the late time expansion of the Universe mainly come from distance measurements via two methods: 1) standard candles, i.e. objects whose intrinsic luminosity can be considered as known, 2) standard rules, i.e. features whose comoving size is known. A typical example of the standard candles are the Type Ia Supernovae (SNIa), which are considered to be standardizable candles, because their astrophysical processes are known and hence their intrinsic luminosity can be inferred. 

However, there is a recent claim that the SNIa might not be good standardizable candles after all. In Ref.~\cite{Kang:2019azh} (see also~\cite{Kim:2019npy,Rigault:2014kaa}), the authors examined whether the absolute magnitude (hereafter $M_{\rm B}$) of SNIa could change with host morphology, host mass, and local star formation rate, finding a significant correlation with the stellar population age. This suggests a redshift evolution of their intrinsic luminosity in a manner that could mimic dark energy, thus challenging the core assumption of the SNIa. Such a result, if validated, could strengthen the claims that the evidence for cosmic acceleration is not robust yet, as discussed in~\cite{Nielsen:2015pga, Dam:2017xqs, Colin:2018ghy} - and disputed by~\cite{Yang:2019fjt, Rubin:2019ywt, Nadathur:2020kvq}. However, other recent analyses reported no further evidence for the SNIa intrinsic luminosity evolution ~\cite{Huang:2020mub, Koo:2020ssl, Rose:2020shp, Brout:2020msh, Kazantzidis:2020tko, DiValentino:2020evt}.  

Therefore, it is crucial to determine whether the SNIa are reliable standard candles or they are affected by yet unresolved systematics. One possible approach would be to directly fit the SNIa absolute magnitude in different redshift bins, so we can examine whether there is any hint at a significant fluctuation of its expected value in any redshift range. A similar idea was pursued in~\cite{Koo:2020ssl}, where the authors performed an MCMC analysis on the SNIa light curve parameters in different redshift bins, and using different background cosmologies - besides a non-parametric approach to reconstruct the observed distance modulus of SN Ia. They obtained consistent results with all dark energy models except for the $M_{\rm B}$, since it shifts the distance moduli by a constant amount and thus cannot be constrained.

In this paper we analyse the cosmological implications of an absolute magnitude  $M_{\rm B}$ which could vary with respect to the redshift. Rather than looking for astrophysical effects or other systematics that may affect $M_{\rm B}$, as in~\cite{Kang:2019azh}, we carry out empirical analyses directly on the SNIa observations to check whether the data favours any significant redshift evolution of $M_{\rm B}$. We consider a plethora of very different approaches such as a Taylor expansion of the absolute magnitude, modified gravity models, non-homogeneous models and lastly, direct data binning. The goal of this work is: if the absolute magnitude was varying with redshift, which cosmological model would be favoured? Anticipating the result, we did not find any evidence towards new physics when the absolute magnitude of SNIa is allowed to be redshift dependent.  

The paper is organised as followed: in Section~\ref{sec:functions} we report the basic equation for the cosmological models that will be tested against the observations, in Section~\ref{sec:models} are listed the different cases, in Section~\ref{sec:results} we present the analysis and the results found, and we conclude in Section~\ref{sec:conclusions}.

\section{Theory}\label{sec:functions}
Here we review the basic equations and notation for the background evolution for the different cosmological models we consider in our analysis.

\subsection{Standard dark energy models}
Assuming general relativitiy (GR) and the  Friedmann-Lema\^itre-Robertson-Walker (FLRW) Universe, the Hubble parameter for matter and dark energy can be written as 
\be
\label{eq:hz}
\frac{H^2(z)}{H_0^2} = \Omega_{\rm m,0}(1+z)^3 + (1-\Omega_{\rm m,0})(1+z)^{3(1+\hat{w}(z))}\,,
\ee
which is valid for a flat Universe and a generic dark energy model with an equation of state:  
\bea
\label{eq:wz}
\hat{w}(z)  &=& \frac{1}{\ln(1+z)}\int_0^z \frac{w(z')}{1+z'}{\rm d}z'\\
w(z)&\equiv& p_{\rm DE}(z)/\rho_{\rm DE}(z) \,.
\eea
where $\rho_{\rm DE}$ and $p_{\rm DE}$ are the density and pressure of dark energy, respectively. If we fix $w=-1$ then we recover the cosmological constant model $\Lambda$CDM, while if we leave $w$ free but constant, we recover the $w$CDM model. 

The absolute magnitude $M_{\rm B}$ is related to the logarithm of the luminosity, while the apparent magnitude $m$ is related to the flux received. Absolute and apparent magnitude are related to the luminosity distance via the relation:
\begin{equation}\label{eq:mu_th}
\mu_{\rm th} = m-M_{\rm B} = 5\log_{10}{D_{\rm L}(z)} + 25,    
\end{equation}
where $\mu_{\rm th}$ is the distance modulus. The luminosity distance $D_{\rm L}(z)$ is connected to the cosmological model and the Hubble parameter via 
\begin{equation}\label{eq:dlz}
D_{\rm L}(z) = \frac{c}{H_0}(1+z)\int_0^z \frac{{\rm d}z'}{h(z')} = \frac{c}{H_0}d_{\rm L}(z)\,,
\end{equation}
where we expressed the Hubble parameter as $H^2(z)=H_0^2\,h^2(z)$ and $d_{\rm L}(z)$ is the dimensionless luminosity distance. 
If the absolute magnitude $M_{\rm B}$ is known, then the luminosity distance can be derived from Eq.~\eqref{eq:mu_th} once the relative magnitude $m$ is measured. However, Eq.~\eqref{eq:mu_th} contains another constant term, i.e. $c/H_0$, which cannot be completely disentangled from the absolute magnitude using the SNIa alone. It is, then, convenient to rewrite Eq.~\eqref{eq:mu_th} in such a way that this constant is absorbed. Doing so, we find $H_0$ is degenerate with $M_{\rm B}$ and we have: 
\be
\label{eq:mu_th_1}
\mathcal{M}=  25 + 5\log_{10}(c/H_0) + M_{\rm B}\,,   
\ee
which is the only parameter that can be directly constrained from observations, rather than $M_{\rm B}$. 

Usually, $\MM$ is assumed to be a constant parameter because all the quantities entering into the above expression are assumed to be constant in GR. However, in this work we will break this assumption and we will allow $\MM$ to vary with the redshift. This dependence might come from a more complex relation between the absolute magnitude $M_{\rm B}$ and the environment of the host galaxy, inhomogeneous cosmological models such as e.g. the  Lema\^itre-Tolman-Bondi (LTB) models, or even modified gravity models, as we will describe in what follows. 

Finally, the theoretical distance modulus is given by 
\be
\mu_{\rm th}(z) = \MM(z) + 5\log_{10} d_L(z)\,, 
\label{eq:th-dist-mod}
\ee
where $d_{L}(z)$ is the dimensionless luminosity distance given in Eq.~\eqref{eq:dlz}, and $\MM(z)$ represents our new modified magnitude relation. 

We often prefer not to make strong assumptions on the absolute magnitude $\MM$ of the SNIa, so it is preferable to marginalise over $\MM(z)$ or at least to consider it as an independent and free parameter, thus we will also report on this in the next sections. 

\subsection{Modified gravity}
In modified gravity models in general   Newton's constant $G_{\textrm{N}}$ can be time and scale dependent $G_{\textrm{N}}\rightarrow G_{\textrm{eff}}(a,k)$, e.g. see Ref.~\cite{Tsujikawa:2007gd,Nesseris:2008mq,Nesseris:2009jf}. This evolution not only affects the large scale (LSS) of the Universe, but also the peak luminosity of SnIa as the latter is proportional to the mass of nickel synthesized which is a fixed fraction of the Chandrasekhar mass $M_{\textrm{Ch}}$ varying as $M_{\textrm{Ch}}\sim G_{\textrm{eff}}^{-3/2}$, see Ref.~\cite{Gaztanaga:2001fh}, where $G_{\textrm{eff}}$ is the effective Newton's constant. Therefore, the SnIa peak luminosity varies like $L\sim G_{\textrm{eff}}^{-3/2}$ and the corresponding SnIa absolute magnitude evolves like \cite{Nesseris:2006jc}
\be
\mathcal{M}(z) = \mathcal{M}_0+\frac{15}{4} \log_{10} \frac{G_{\textrm{eff}}}{G_\textrm{N}},\label{eq:mgeff}
\ee
where $G_\textrm{N}$ is the bare Newton's constant as measured in a Cavendish-type experiment and $\mathcal{M}_0$ is the absolute magnitude in GR. In the presence of an evolving Newton's constant, the luminosity distance also picks up a correction compared to GR, due to the modified Friedmann equation. In order to model this effect in a general manner, we follow Ref.~\cite{Nesseris:2006jc} and write the luminosity distance as:
\be
D_{\rm L}(z) = \frac{c}{H_0}(1+z)\int_0^z dz'\sqrt{\frac{G_\textrm{N}}{G_{\textrm{eff}}(z')}}\frac{1}{h(z')}.\label{eq:dLG} 
\ee
Hence, with this phenomenological approach we can now imitate a modified gravity model without loss of generality. Similarly, quantities that depend on the Hubble parameter, such as the sound horizon 
\be
r_s(z_{*})=\int_0^{a_{*}}\frac{c_s}{a^2 H(a)}{\rm d}a\nonumber
\ee will be rescaled by the same factor, i.e. 
\be 
H(a)\rightarrow H(a) \frac{G_{\textrm{eff}}(z')}{G_\textrm{N}}\,.\nonumber
\ee
In what follows, we will consider two variations of the $G_{\textrm{eff}}$ parametrization of Ref.~\cite{Nesseris:2017vor}. 
First we consider the form
\be
\frac{G_{\textrm{eff}}(z)}{G_\textrm{N}}=1+g_{\rm a} \left(\frac{z}{1+z}\right)^n-g_{\rm a} \left(\frac{z}{1+z}\right)^{2n},\label{eq:Geff}
\ee
which asymptotes to unity at early and late times, thus recovering the standard value of the Newton's constant. We also consider the form 
\be
\frac{G_{\textrm{eff}}(z)}{G_\textrm{N}}=1+g_{\rm a} \left(\frac{z}{1+z}\right)^n,\label{eq:Geff1}
\ee
which asymptotes to unity at late times, but at early times it has the limit $G_{\textrm{eff}}(z\gg1)/G_\textrm{N}=1+g_{\rm a}$, which implies that $g_{\rm a}$ has to be well within the range $g_{\rm a}\in[-0.1,0.1]$ in order to be consistent with Big Bang Nucleosynthesis (BBN) constraints \cite{Bambi:2005fi}. 

However, in some modified gravity theories, e.g. in the well-known case of $f(R)$, the  scalar degree of freedom may be massive in high densities and small scales, such as the characteristic scale of a white dwarf. Then the scalar field becomes so heavy that the effects of the modified gravity model are  screened, thus becoming undetectable. In a nutshell, this describes the chameleon mechanism, see for example Ref.~\cite{Brax:2008hh} and references therein. 

As a result, this implies that for phenomena inside the screened regime, any possible correction to the absolute magnitude from the modified gravity 
can be neglected since $G_{\textrm{eff}}\simeq G_{\rm N}$. However, the effect of a modified Newton's constant will be essential for  models where a chameleon is absent and here we focus in the latter case \cite{Burrage:2017qrf}.

\subsection{Cosmic voids and the LTB model}
Here we briefly describe the Lemaitre-Tolman-Bondi (LTB) models, which describe a local void. One main difference with the traditional $\Lambda$CDM model and the FLRW metric is that most quantities, not only depend on the cosmic time, but also on the radial distance $r$. Specifically, these models can be fully defined by the matter density $\Omega_{\rm m}(r)$ and the Hubble expansion rate $H(r)$, for which we use the constrained GBH parametrization of Ref.~\cite{GarciaBellido:2008nz}
\bea
\Omega_{\rm m}(r)&=& \left(\Omega_{\textrm{in}}-\Omega_{\textrm{out}}\right) \frac{1-\tanh\left[(r-r_0)/2\Delta r\right]}{1+\tanh\left[r_0/2\Delta r\right]}+\nonumber \\
&+&\Omega_{\textrm{out}}\,, \label{eq:ltb1} \\
H_0(r)&=& H_0 \left(\frac{1}{\Omega_{\rm k}(r)}-\right.\nonumber \\
 &-&\left.\frac{\Omega_{\rm m}(r)}{\sqrt{\Omega_{\rm k}(r)^3}}\sinh^{-1}\sqrt{\frac{\Omega_{\rm k}(r)}{\Omega_{\rm m}(r)}}\right)\,,\label{eq:ltb2}
\eea 
where $\Omega_{\rm k}(r)=1-\Omega_{\rm m}(r)$, $\Omega_{\textrm{out}}$ is the value of the matter density at infinity, $\Omega_{\textrm{in}}$ is the value of the matter density at the center of the void, $r_0$ is the size of the void, while $\Delta r$ is a scale that characterises the transition to uniformity. While the LTB model predicts no intrinsic evolution of the absolute magnitude, we also consider it in order to test for any effects of voids in the determination of  $\MM$.\\

\section{Evolution of the absolute magnitude\label{sec:models}}
In order to test for a possibly evolving absolute magnitude $\MM$, we will consider a plethora of different scenarios, ranging from Taylor expansions of $\MM(z)$, to binning, the effects of modified gravity or a cosmic void as parametrized via the LTB model.

First, we start with the expansions around $z=0$, for which we assume that the absolute magnitude is dependent on redshift through 
\begin{equation}
    \MM(z) = M_0 +M_1\cdot f(z), 
    \label{eq:abs_magn}
\end{equation}
where the function $f(z)$ needs to be defined. In what follows, we consider the four different phenomenological scenarios for $f(z)$:
\begin{itemize} 
\item {\bf Model 1} $f(z) = z/(4 z + 1)$, a CPL model that does not grow as fast at high redshifts in order to match the Hubble residual diagram in \cite{Kang:2019azh},
\item {\bf Model 2} $f(z) = z$, a first order Taylor expansion around $z=0$, in order to test deviations at small redshifts,
\item {\bf Model 3} $f(z) = \left(z/(1+z)\right)^\alpha$, a generalized CPL-like model, where the coefficient $\alpha$ controls the growth of the absolute magnitude, 
\item {\bf Model 4} $f(z) = \exp({\alpha z})$ with $\alpha$ a free parameter, motivated by the evolution model of Ref.~\cite{Kang:2019azh} and references there-in.
\end{itemize} 
It is clear that the phenomenological models reported above are meant to capture possible astrophysical effects or systematics in the data. However, the idea is to consider these variation on the absolute magnitude and study their implications to the cosmological analysis.

Our second approach is to consider the modification of the absolute magnitude due to modification of gravity. 
In particular, we consider two different scenarios based on generic modifications of GR and an evolving Newton's constant, where the luminosity distance is modified in accordance with Eq.~\eqref{eq:dLG}, while the absolute magnitude has an extra correction as in Eq.~\eqref{eq:mgeff}. Specifically, we consider the models:
\begin{itemize}
\item {\bf Model 5} $G_{\rm eff}(z)$ given by Eq.~\eqref{eq:Geff}.
\item {\bf Model 6} $G_{\rm eff}(z)$ given by Eq.~\eqref{eq:Geff1}.
\end{itemize}
We also consider the effects of a void centered at us, as modeled by the LTB model
\begin{itemize}
\item {\bf Model 7} The LTB using the GBH parameterization for the matter density profile $\Omega_m(r)$ given by the GBH model of Eq.~\ref{eq:ltb1}, see Ref.~\cite{GarciaBellido:2008nz}.
\end{itemize}
Finally, as a last model independent method, we bin directly the data, using two different schemes: 
\begin{itemize}
\item {\bf Binning scheme 1}: we assume $\Omega_{\rm m,0}$ constant in redshift and we bin only $\MM_i$.
\item {\bf Binning scheme 2}: we bin pairs of $\Omega_{\rm m,0}-\MM_i$. 
\end{itemize}
In the following sections we will describe the data we use and present our constraints of the aforementioned models.

\section{Analysis and results}\label{sec:results}
Here we discuss our methodology for analysis the SnIa data and the result we obtained for the models mentioned in Sec.~\ref{sec:models}. In particular, we will present the explicit likelihoods used in the minimization or the particular binning techniques.

The data used in this work is the  Pantheon catalog which contains 1048 points \cite{Scolnic:2017caz} given as
\begin{equation}\label{eq:mu_obs}
\mu_{\rm obs} = m^\star_{\rm B} - (M_{\rm B} - \alpha X_1 - \beta C)\,,  
\end{equation}
which correspond to light-curve parameters, the observed B-band peak magnitude ($m^\star_{\rm B}$),
the stretching of the light curve ($X_1$), and the SN color at maximum brightness ($C$). $M_{\rm B}$, $\alpha$, and $\beta$ are hyperparameters that are marginalised over when performing cosmological model inferences, i.e., nuisance parameters. 

Finally, it should be noted that the data as we use them here in the form of the distance modulus, have been derived making mild model assumptions \cite{Scolnic:2017caz}. A more complete analysis would be to consider directly the light of curves of the SNIa, which contain information on the astrophysical parameters of the stretch and the color, and perform the same analysis. However, since we would like to avoid mixing the astrophysical effects, as codified by the stretch and the color, we prefer to use the version of the data where these have been marginalized over.

\subsection{Varying absolute magnitude}
Our analysis relies on the likelihood, so we need to adapt our usual statistical analysis to the variable absolute magnitude models, namely models 1-4. In this case, we make use of Eq.~\eqref{eq:abs_magn}, then the $\chi^2$ can be written as 
\begin{equation}
    \chi^2 =  D^{T} C^{-1} D
    \label{eq:chi2_new}
\end{equation}
being $D = \mu_i -\mu_{\rm th}(z_i)$ the data vector, $\mu_i$ are the data in Eqs.~\eqref{eq:mu_obs} and $\mu_{\rm th}$ is given by Eq.~\eqref{eq:th-dist-mod}. Using index notation we can write the data vector as 
\be
\DD^i = \mu^i-5\log_{10} d_L(z^i)\,,\nonumber
\ee
where the superscript refers to the redshift of the data; then Eq.~\eqref{eq:chi2_new} reads 
\be
\chi^2 = \left(\DD^i - \MM(z^i)\right) C^{-1}_{ij}\left(\DD^j - \MM(z^j)\right)\,.
\label{eq:chi2_new_2}
\ee
The quantity $\MM(z^i)$ is the varying absolute magnitude given in Eq.~\eqref{eq:abs_magn} which contains both $\MM_0$ and $\MM_1$. 

We can also marginalize over $\MM_0$ and $\MM_1$, see also Appendix~\ref{sec:appendix1}, and the $\chi^2$ then becomes:
\bea
\chi^2_{\rm marg,\,\MM} &=& \ln \left(A D-H^2\right) +\nonumber \\
&+&\frac{A F^2+B^2 D-2 B F H}{H^2-A D} +\nonumber \\
&+& E-2 \log (2 \pi )\,,
\eea
where the terms in the $\chi^2_{\rm marg,\,\MM}$ are given by
\bea
A &=& I^i C^{-1}_{ij}I^j\nonumber\\
B &=& \DD^i C^{-1}_{ij}I^j \nonumber\\
D &=& f(z^i) C^{-1}_{ij}f(z^j)\nonumber\\
E &=& \DD^i C^{-1}_{ij}\DD^j \nonumber\\
F &=& \DD^i C^{-1}_{ij}f(z^j) \nonumber\\
H &=& I^i C^{-1}_{ij}f(z^j)\,.\nonumber
\eea
In the above expressions, the data must satisfy $A D-H^2>0$ in order for the argument of the logarithm to be positive, which we find that for the Pantheon data is always true. For a more detailed derivation of the marginalized $\chi^2$ we refer the interested reader to Appendix \ref{sec:appendix1}.

\begin{table}[htp]
\begin{center}
  \begin{tabular}{ |ccccc| }
    \hline
 & $\Omega_{\rm m,0}$ &$w$ & $\alpha$ &   $\chi^2_{\rm min}$\\
  \hline  
  \hline
$\Lambda{\rm CDM}$ &  $ 0.299 \pm 0.020 $ & $-1$ & - & 1034.7\\
{\bf M1} &  $ 0.308 \pm 0.059 $ & $-1$ & - & 1036.9\\
{\bf M2} &  $ 0.032 \pm 0.100$ & $-1$ & - & 1039.2\\
{\bf M3} &  $ 0.242\pm 0.230 $ & $-1$ & $ 1.758\pm 0.386$ & 1037.5\\
{\bf M4} &  $ 0.285\pm  0.024$ & $-1$ & $1.638\pm 0.175$ &  1043.1\\
\hline
\hline
$w$CDM &  $ 0.314 \pm 0.022 $ & $-1.046\pm 0.054$ & - & 1034.7\\
{\bf M1} &  $0.312\pm 0.051$ & $-1.030 \pm 0.109$ &  - &1036.9\\
{\bf M2} &  $0.079 \pm 0.092$ & $-0.944 \pm 0.077$ & - & 1039.3\\
{\bf M3} &  $ 0.214\pm 0.083 $ & $ -0.848\pm 0.142 $ & $6.173 \pm 0.336$& 1031.4\\
{\bf M4} &  $ 0.173\pm 0.108$ & $ -0.882\pm 0.185$ & $0.807\pm 0.165$ &  1040.4 \\
\hline
\end{tabular} 
\end{center}
\caption{The best fit values for model 1 ({\bf M1}), model 2 ({\bf M2}), model 3 ({\bf M3}) and model 4 ({\bf M4}) marginalized over the absolute magnitudes $\MM_0$ and $\MM_1$. The results have been obtained assuming flatness.}
\label{tab:bf-models}
\end{table}
%
%%%%%%%%%%%%%%
%%%%%% THIS IS THE ORIGINAL TABLE %%%%%%%%%%%%%%%
%\begin{table*}[t]
%\begin{center}
%  \begin{tabular}{ |cccccc| }
%    \hline
% & $\MM_0$ & $\MM_1$ & $\Omega_{\rm m,0}$ & $\alpha$ &   $\chi^2_{\rm min}$\\
%  \hline  
%  \hline
%{\bf $\Lambda$CDM} &  $ -1.191 \pm 0.011$ &  $ 0 $ &  $ 0.299 \pm 0.022 $ & - & 1025.63 \\
%{\bf M1} &  $ -1.194 \pm 0.020 $ &  $ 0.061\pm 0.381 $ &  $ 0.299 \pm 0.064$ & - & 1025.60 \\
%{\bf M2} &  $ -1.190 \pm 0.013$ &  $ -0.474\pm 0.177 $ &  $ 0.032 \pm 0.069$ & - & 1024.89 \\
%{\bf M3} &  $ -1.192 \pm 0.016 $ &  $ 0.031\pm 0.462 $ &  $ 0.311 \pm0.188 $ & 1 & 1025.62 \\
%{\bf M3}${}_{\textrm{MD}}$ &  $-1.195 \pm 0.012$ &  $ 1.204\pm 0.055 $ &  1 & 1 & 1025.93 \\
%{\bf M4} &  $ -1.193 \pm 0.570 $ &  $ 0.001\pm 0.579 $ &  $ 0.298 \pm 0.287 $ & -1 & 1025.63 \\
%\hline
%\end{tabular} 
%\end{center}
%\caption{The best fit values for model 1 ({\bf M1}), model 2({\bf M2}), model 3 ({\bf M3}) and model 4 ({\bf M4}) allowing the absolute magnitudes $\MM_0$ and $\MM_1$ free. The results have been obtained assuming flatness.\snc{Make single column}}
%\label{tab:bf-models1}
%\end{table*}

%\begin{table*}[t]
\begin{table}[htp]
\begin{center}
  \begin{tabular}{ |cccccc| }
    \hline
 & $\MM_0$ & $\MM_1$ & $\Omega_{\rm m,0}$ & $\alpha$ &   $\chi^2_{\rm min}$\\
  \hline  
  \hline
${\bf \Lambda}$ &  $ -1.191 \pm 0.011$ &  $ 0 $ &  $ 0.299 \pm 0.022 $ & - & 1025.6 \\
{\bf M1} &  $ -1.194 \pm 0.020 $ &  $ 0.061\pm 0.381 $ &  $ 0.299 \pm 0.064$ & - & 1025.6 \\
{\bf M2} &  $ -1.190 \pm 0.013$ &  $ -0.474\pm 0.177 $ &  $ 0.032 \pm 0.069$ & - & 1024.9 \\
{\bf M3} &  $ -1.192 \pm 0.016 $ &  $ 0.031\pm 0.462 $ &  $ 0.311 \pm0.188 $ & 1 & 1025.6 \\
${\bf \overline{M3}}$ &  $-1.195 \pm 0.012$ &  $ 1.204\pm 0.055 $ &  1 & 1 & 1025.9 \\
{\bf M4} &  $ -1.193 \pm 0.570 $ &  $ 0.001\pm 0.579 $ &  $ 0.298 \pm 0.287 $ & -1 & 1025.6 \\
\hline
\end{tabular} 
\end{center}
\caption{The best fit values for model 1 ({\bf M1}), model 2 ({\bf M2}), model 3 ({\bf M3}) and model 4 ({\bf M4}) allowing the absolute magnitudes $\MM_0$ and $\MM_1$ free. The model 3 (${\bf \overline{M3}}$) refers to {\bf M3} where we fix $\Omega_{\rm m,0} = 1$. The results have been obtained assuming flatness and fixing the dark energy equation of state $w=-1$. ${\bf \Lambda}$ refers to the $\Lambda$CDM model. }
\label{tab:bf-models1}
\end{table}

In Tab.~\ref{tab:bf-models} we report the results for the models 1 to 4 marginalizing over $\MM_0-\MM_1$. The upper part of the table shows the results when we fix the dark energy equation of state parameter to its cosmological constant value $-1$, whereas in the lower part we allow also $w$ to vary. In general, the analysis does not point to a favoured model as the $\chi^2$ increases when the model becomes more complex: model 2 and model 4, representing a linear and an exponential growth of the absolute magnitude, are rather disfavoured, having very large values of the $\chi^2$. The only exception is model 3 which has $\chi^2_{\rm min} = 1031.45$. However, the best fit values for model 3 are still comparable with the $\Lambda$CDM model within the $1\sigma$ errors. Model 3 corresponds to a generalized CPL parameterization $(z/(1+z))^\alpha$, hence a high values of $\alpha$, like the one found $\alpha \approx 6$, prevents the absolute magnitude to grow too fast.

We also performed the analysis allowing the absolute magnitude parameters $\MM_0$ and $\MM_1$ to vary together with the other parameters. Even in this case we do not find any deviation from the $\MM(z) = $ const. The results are reported in Table~\ref{tab:bf-models1}. Note that in this case, the values of the $\chi^2$ overall are smaller, due to the different normalization, as we are not marginalizing over $\MM_0-\MM_1$.

\begin{table}[htp]
\begin{center}
  \begin{tabular}{ |ccccc| }
    \hline
 & $\Omega_{\rm m,0}$ & $g_{\rm a}$ & $n$ &   $\chi^2_{\rm min}$\\
  \hline  
  \hline
${\bf \Lambda}$  & $ 0.299 \pm 0.022 $ & 0 & 0 & 1034.73 \\
{\bf M5} &  $ 0.344 \pm 0.155 $ &  $ 0.257 \pm 0.858 $ &  $  2 $ & 1034.65 \\
{\bf M6}${}_\textrm{MD}$ &  $1$ &  $1.676\pm0.141$ &  $1.259\pm0.133$ & 1035.58 \\
{\bf M6} &  $0.270\pm0.124$ &  $-0.130\pm0.554$ &  $2$ & 1034.68 \\
\hline
\end{tabular} 
\end{center}
\caption{Constraints on the modified gravity models M5 and M6 given by the $G_{\textrm{eff}}$ expressions of Eqs.~\eqref{eq:Geff} and \eqref{eq:Geff1}, marginalizing over $\mathcal{M}_0$ and assuming flatness. ${\bf \Lambda}$ refers to the $\Lambda$CDM model.}
\label{tab:bf-models2}
\end{table}

\begin{table}[b]
\begin{center}
  \begin{tabular}{ |cccc| }
    \hline
 $\Omega_{\rm m,in}$ & $r~[\textrm{Gpc}]$ & $\Delta r~ [\textrm{Gpc}]$ &   $\chi^2_{\rm min}$\\
  \hline  
  \hline  $  0.298\pm 0.007 $ &  $1.0$ &  $0.30$ & 1083.33   \\  $  0.197\pm 0.008 $ &  $1.5$ &  $0.45$ & 1046.16 \\
  $  0.156\pm 0.005 $ &  $1.8$ &  $0.54$ & 1041.43 \\
  $  0.200\pm 0.004 $ &  $2.0$ &  $0.60$ &  1049.99\\
\hline
\end{tabular} 
\end{center}
\caption{The best-fit parameters assuming the LTB model with a void centered at us, marginalizing over $\mathcal{M}_0$. The parameters $r$ and $\Delta r$ are the size and scale of  transition to uniformity of the void, $\Omega_{\rm m,in}$ is the matter density at the center, while $\Omega_{\rm m,out}=1$. The comoving void sizes of $r=(1.0,1.5,1.8,2.0)\textrm{Gpc}$ correspond to redshifts $z=(0.24, 0.37, 0.45, 0.51)$ in Planck 2018 $\Lambda$CDM.\label{tab:bf-models3}}
\end{table}

\begin{figure*}[t!]
\centering
\includegraphics[width=0.4\textwidth]{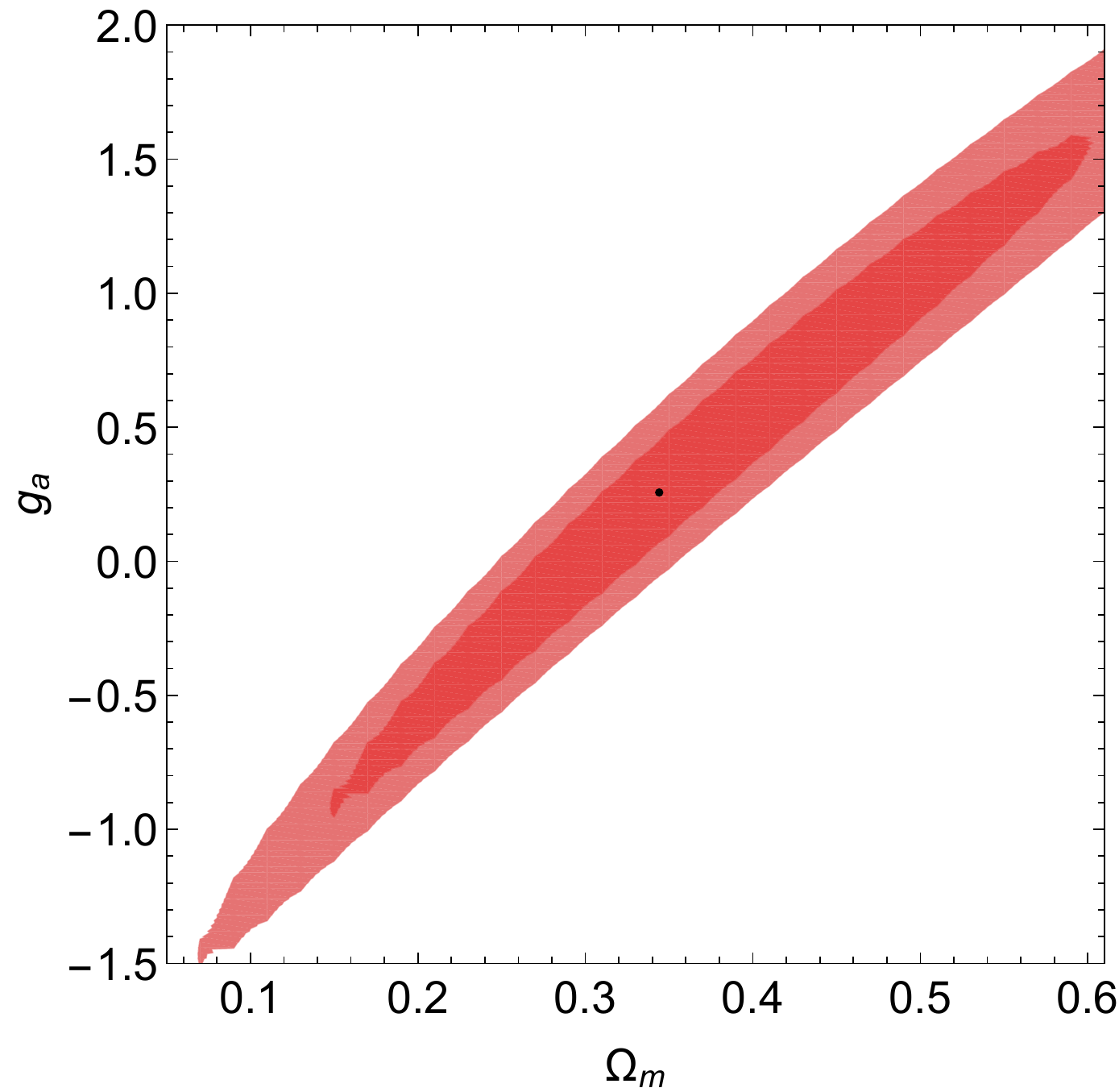}
\includegraphics[width=0.4\textwidth]{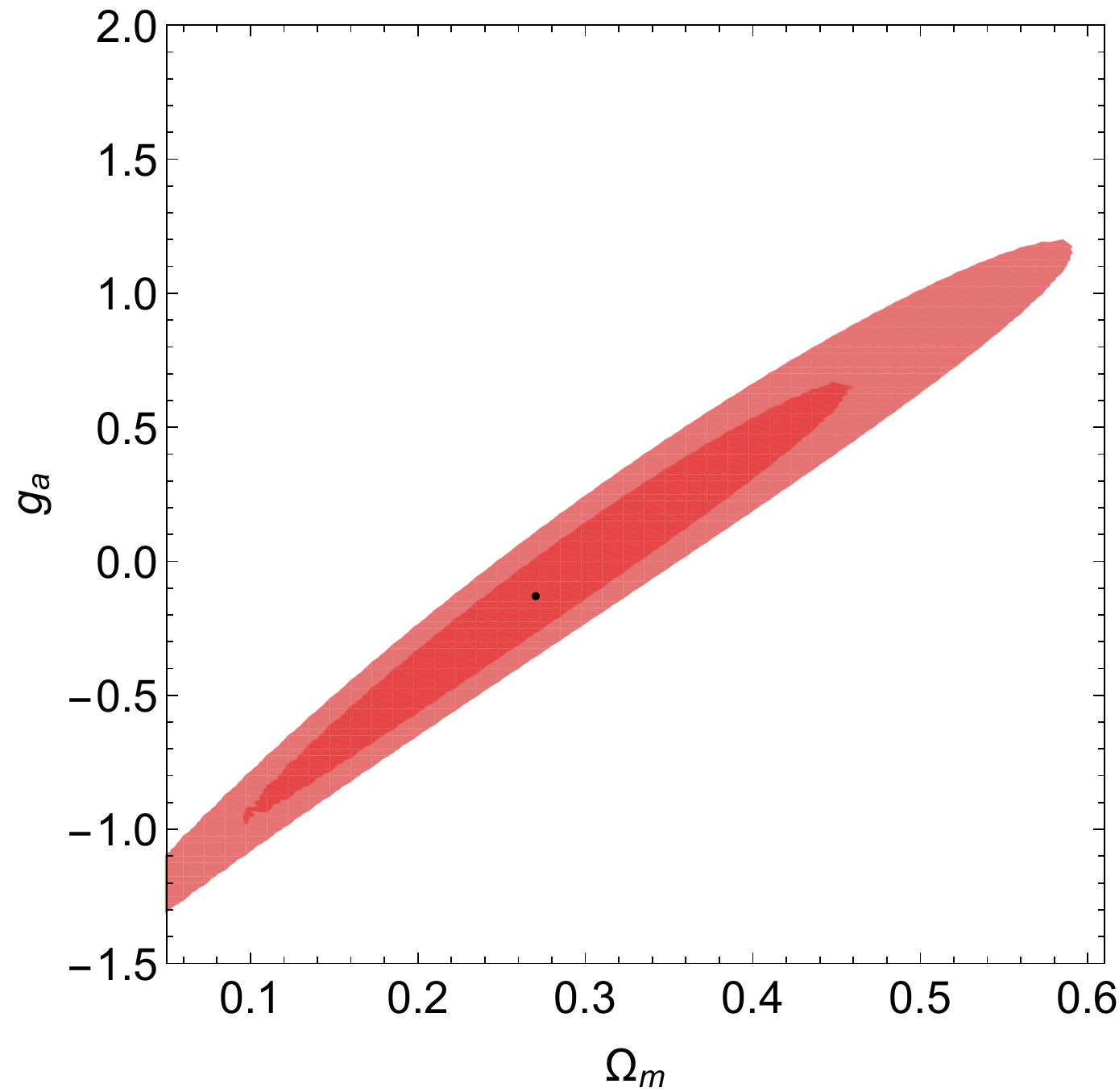}
\caption{Contours for {\bf Model 5} (left) and {\bf Model 6} (right) for the $G_{\textrm{eff}}$ models given by Eqs.~\eqref{eq:Geff} and \eqref{eq:Geff1} for $n=2$, assuming SnIa only.}
\label{fig:geff_marge}
\end{figure*}

\begin{figure}[t!]
\centering
\includegraphics[width=0.45\textwidth]{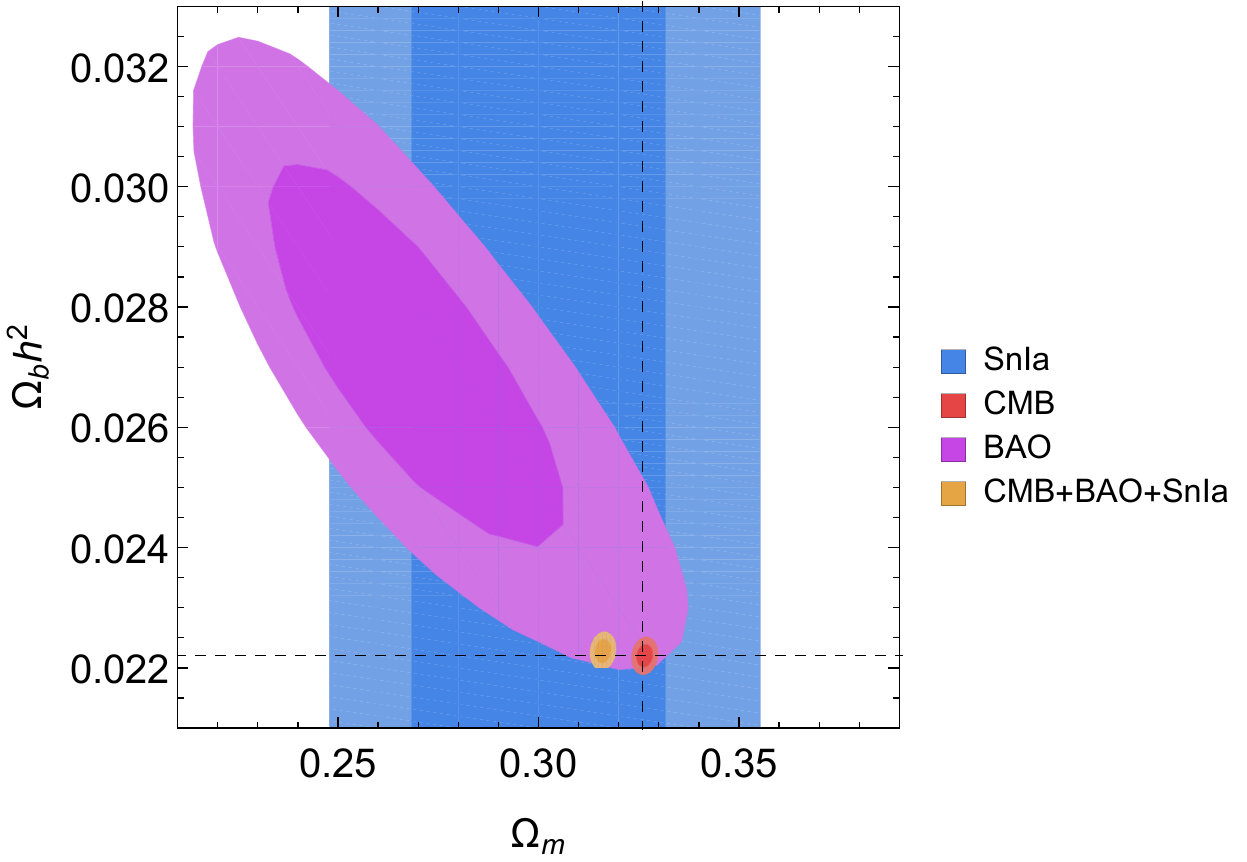}
\caption{Confidence contours in the $(\Omega_{\rm m,0},\Omega_{\rm b}h^2)$ parameter plane for $\Lambda$CDM assuming SnIa+CMB+BAO. The dashed lines correspond Planck.}
\label{fig:geff_marge1}
\end{figure}

\begin{figure}[h!]
\centering
\includegraphics[width=0.45\textwidth]{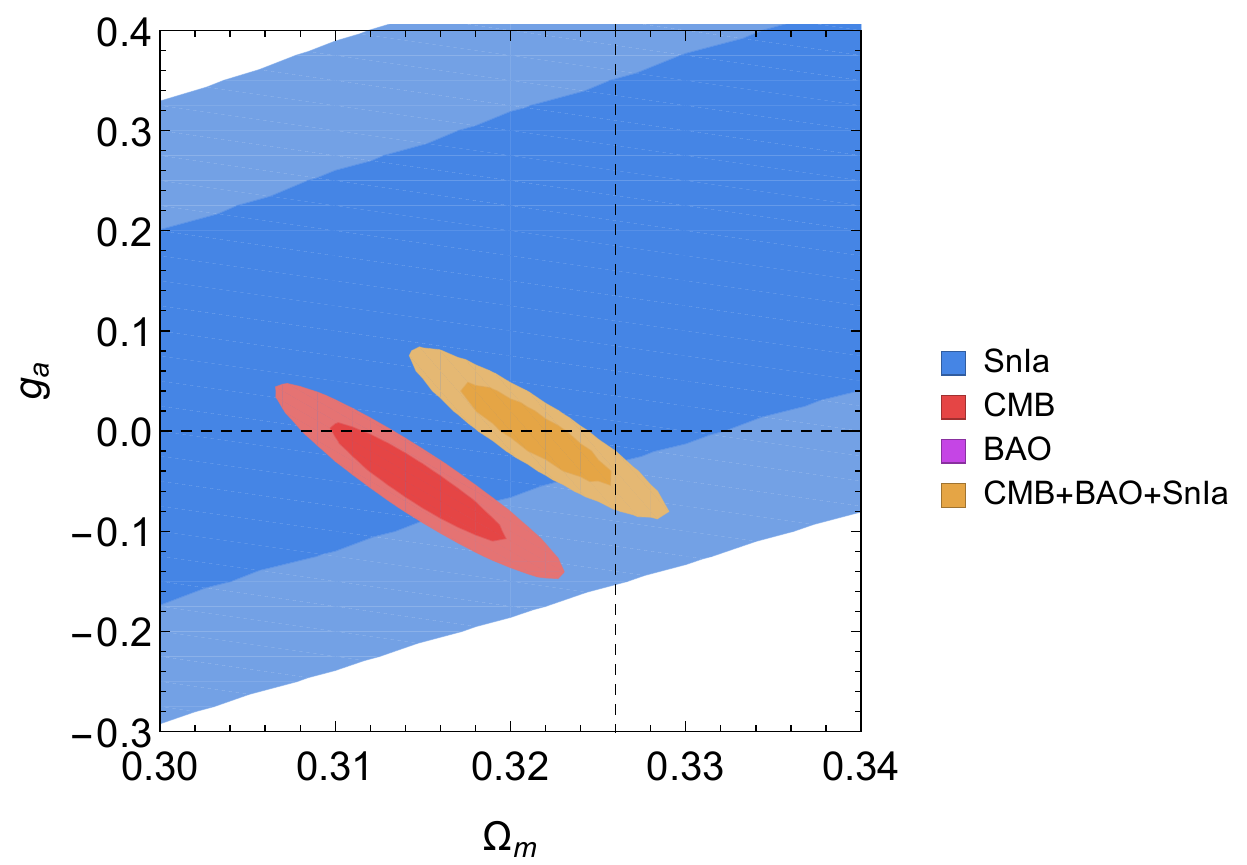}
\caption{Confidence contours in the $(\Omega_{\rm m,0},g_a)$ parameter plane for {\bf Model 5} for the $G_{\textrm{eff}}$ model given by Eqs.~\eqref{eq:Geff}for $n=2$, assuming SnIa+CMB+BAO. The dashed lines correspond to GR and Planck. The BAO contour is outside the plot range show here.}
\label{fig:geff_marge2}
\end{figure}

\subsection{Modified gravity}
Here we discuss the results obtained from the modified gravity model of Eqs.~\eqref{eq:Geff} and \eqref{eq:Geff1}, which we denoted a models 5 and 6 respectively. In particular, we show the best-fit values for $\Omega_{\rm m,0}$, $g_{\rm a}$ and $n$ in Table~\ref{tab:bf-models2}. We consider the cases where both $\Omega_{\rm m,0}$ and $g_{\rm a}$ are free to vary, but also when  $\Omega_{\rm m,0}$ is fixed to unity, thus corresponding to a matter dominated model (denoted by $M6_\textrm{MD}$ in the table). In Fig.~\ref{fig:geff_marge} we also show the $1\sigma$ and $2\sigma$ confidence contours for the two models: Model 5 (left) and Model 6 (right) panels respectively for $n=2$. In both cases, the contours are in agreement with $g_a=0$ and GR within the errors. 

As seen, in  Table~\ref{tab:bf-models2} the analysis finds no deviation from GR or any preference over the \lcdm model in terms of the $\chi^2$, even though Model 5 slightly outperforms \lcdm by $\delta \chi^2\simeq 0.1$, but at the same time has one more free parameter. 

In this regard, we also consider the combination of CMB+SnIa+BAO data, in order to examine if these additional data sets can provide any further constraining power. We show the resulting contours in Fig.~\ref{fig:geff_marge1} for the \lcdm model and in Fig.~\ref{fig:geff_marge2} for Model 5. We show explicitly the contours for the different data sets, in order to highlight their individual constraining power and how each of them breaks the degeneracies. As seen in Fig.~\ref{fig:geff_marge2}, the total combination of CMB+BAO+SnIa (yellow contour), helps break the degeneracies of the SnIa data and leads to much tighter constraints with respect to those of Fig.~\ref{fig:geff_marge}, albeit still consistent with GR.

\subsection{LTB model}
Here we present the constraints on the LTB model with a void centered at us and described by the profiles given by Eqs.~\eqref{eq:ltb1} and \eqref{eq:ltb2}. The best-fit parameters are given in Table~\ref{tab:bf-models3}, where we marginalize over $\mathcal{M}_0$, the parameters $r$ and $\Delta r$ are the size and scale of  transition to uniformity of the void, $\Omega_{\rm m,in}$ is the matter density at the center, while $\Omega_{\rm m,out}=1$. We also assume that the comoving void has a size of $r=(1.0,1.5,1.8,2.0)\textrm{Gpc}$ that corresponds to redshifts $z=(0.24, 0.37, 0.45, 0.51)$ in Planck 2018 $\Lambda$CDM. In this case we find that the LTB model performs significantly worse than the \lcdm model with a $\delta \chi^2 \in [7,50]$, thus not being favored by the data. Note however, that these results depend strongly on the choice of the particular GBH profile given by Eqs.~\eqref{eq:ltb1} and \eqref{eq:ltb2}. While this profile is the most used in literature, some other choice might in principle lead to improved constraints.

\subsection{Binning the data}

\begin{table}[htp]
%\small
\begin{center}
  \begin{tabular}{ |ccccc| }
    \hline
& 2 Bins & 3 Bins &  4 Bins & 5 Bins  \\
  \hline  
  \hline
$\Omega_{\rm m,0}$&  $ 0.29 \pm 0.03 $ & $ 0.3 \pm 0.04 $ &$ 0.32 \pm 0.05 $  & $ 0.36 \pm 0.08 $     \\
$\MM_1$ & $ -1.19 \pm 0.01 $& $ -1.19 \pm 0.01 $& $ -1.19 \pm 0.01 $& $ -1.2 \pm 0.02 $   \\
$\MM_2$ & $ -1.19 \pm 0.02 $ & $ -1.19 \pm 0.02 $& $ -1.18 \pm 0.02 $& $-1.16 \pm 0.02 $    \\
$\MM_3$ &-  & $ -1.18 \pm 0.03 $& $ -1.18 \pm 0.03 $ &  $ -1.18 \pm 0.03 $ \\
$\MM_4$ &- & -& $ -1.17 \pm 0.05 $& $ -1.14 \pm 0.04 $  \\
$\MM_5$ &- & - & -& $ -1.13 \pm 0.07 $  \\
\hline
\hline
$\chi^2_{\rm min}$& 1025.32 & 1024.91 & 1024.41 & 1017.16 \\
\hline
\end{tabular} 
\end{center}
\caption{Constraints on the binned absolute magnitude $\mathcal{M}$ for the first binning strategy. We assume flat $\Lambda$CDM model. }
\label{tab:bf-flat-binned}
\end{table}

\begin{table}[htp]
\begin{center}
  \begin{tabular}{ |ccccc| }
    \hline
&  2 Bins & 3 Bins &  4 Bins & 5 Bins \\
  \hline  
  \hline
$\Omega_{\rm m,0}$ & $ 0.32 \pm 0.07 $ & $ 0.23 \pm 0.11 $ &$ 0.06 \pm 0.1 $ & $ 0.04 \pm 0.19 $  \\
$\MM_1$  & $ -1.19 \pm 0.02 $  & $ -1.2 \pm 0.02 $   & $ -1.22 \pm 0.02 $ & $ -1.23 \pm 0.02 $     \\
$\Omega_{\rm m,0}$ & $ 0.29 \pm 0.04 $& $ 0.34 \pm 0.1 $ & $ 0.53 \pm 0.18 $& $ 0.58 \pm 0.15 $    \\
$\MM_2$   & $ -1.2 \pm 0.03 $  & $ -1.18 \pm 0.04 $  & $ -1.12 \pm 0.05 $ & $ -1.11 \pm 0.03 $  \\ 
$\Omega_{\rm m,0}$ &-& $ 0.32 \pm 0.05 $& $ 0.16 \pm 0.11 $  & $ 0.9 \pm 0.15 $   \\
$\MM_3$         &- & $ -1.17 \pm 0.04 $& $ -1.26 \pm 0.06 $& $ -1.0 \pm 0.05 $   \\
$\Omega_{\rm m,0}$ &-&-& $ 0.33 \pm 0.06 $ & $ 0.34 \pm 0.14 $   \\
$\MM_4$         &-&-& $ -1.15 \pm 0.05 $& $ -1.15 \pm 0.08 $   \\
$\Omega_{\rm m,0}$ &-&-& -& $ 0.3 \pm 0.08 $  \\
$\MM_5$   &-&-&- & $ -1.18 \pm 0.07 $   \\
\hline
\hline
$\chi^2_{\rm min}$ &1025.13 & 1024.6 & 1019.52 & 1011.32 \\
\hline
\end{tabular} 
\end{center}
\caption{Constraints on the binned absolute magnitude $\mathcal{M}$ for the second binning scheme. We assume flat $\Lambda$CDM model.}
\label{tab:bf-flat-binned_mu_om}
\end{table}

A different and  less model-dependent test consists of dividing the data in different redshift bins and considering each bin as an independent data set. In order to maximise the information in each redshift bin, we decide to adopt the strategy of dividing the full set of SNIa data in bins with equal number of points. We chose 4 different binning cases, with $N=\{2, 3, 4, 5\}$ bins. In detail, the edges of the bins are (approximated to the second digit):
\bea
z_{\rm bin} &=& \{0.01,\,0.25,\,2.26\}, \nonumber\,\\
z_{\rm bin} &=& \{0.01,\,0.18,\,0.34,\,2.26\},\nonumber\,\\
z_{\rm bin} &=& \{0.01,\,0.13,\,0.25,\,0.42,\,2.26\},\nonumber\,\\
z_{\rm bin} &=& \{0.01,\,0.10,\,0.20,\,0.30,\,0.51,\,2.26\}\nonumber\,.
\eea
As mentioned in Sec.~\ref{sec:models}, we consider two different schemes: the first one consists in assuming that only the absolute magnitude varies with $z$ and that the matter density $\Omega_{\rm m,0}$ is constant for all the redshifts; for the second scheme, we assume that both $\Omega_{\rm m,0}$ and $\MM$ vary with respect to the redshift $z$. 

The first scheme is more conservative because we assume that the underlying cosmological model is still $\Lambda$CDM and the absolute magnitude may vary with redshift due to, primarily, astrophysical processes; the second binning scheme, instead, has the potential to break the assumption of homogeneity of the Universe and associate the change of the absolute magnitude of the SNIa to a geometric effect by changing the matter density at the same time. 

In tables \ref{tab:bf-flat-binned} and \ref{tab:bf-flat-binned_mu_om} we report the results for the first and second binning scheme, respectively. For both analyses we assumed a flat $\Lambda$CDM model where we fix the dark energy equation of state parameter  $w=-1$. In order to make a clear comparison we report the best fit values in the standard $\Lambda$CDM scenario in Table~\ref{tab:bf-models1}: 
\bea
\Omega_{\rm m, 0} &=& 0.29\pm 0.02\,,\nonumber \\
\MM &=&-1.19 \pm 0.01 \nonumber\,, \\
\chi^2_{\rm min} &=&  1025.63\,.\nonumber 
\eea
For the first binning scheme, the value of $\Omega_{\rm m,0}$ does not change appreciably within $1\sigma$ errors; for instance, the best fit value of $\Omega_{\rm m,0}$ in the 2 bins case is $0.29\pm 0.03$, whereas for the 5 bins case $\Omega_{\rm m0}=0.36\pm 0.08$. The absolute magnitudes $\MM_i$ do not manifest any substantial change when we increase the number of bins.

\begin{figure}[h!]
\centering
\includegraphics[width=0.45\textwidth]{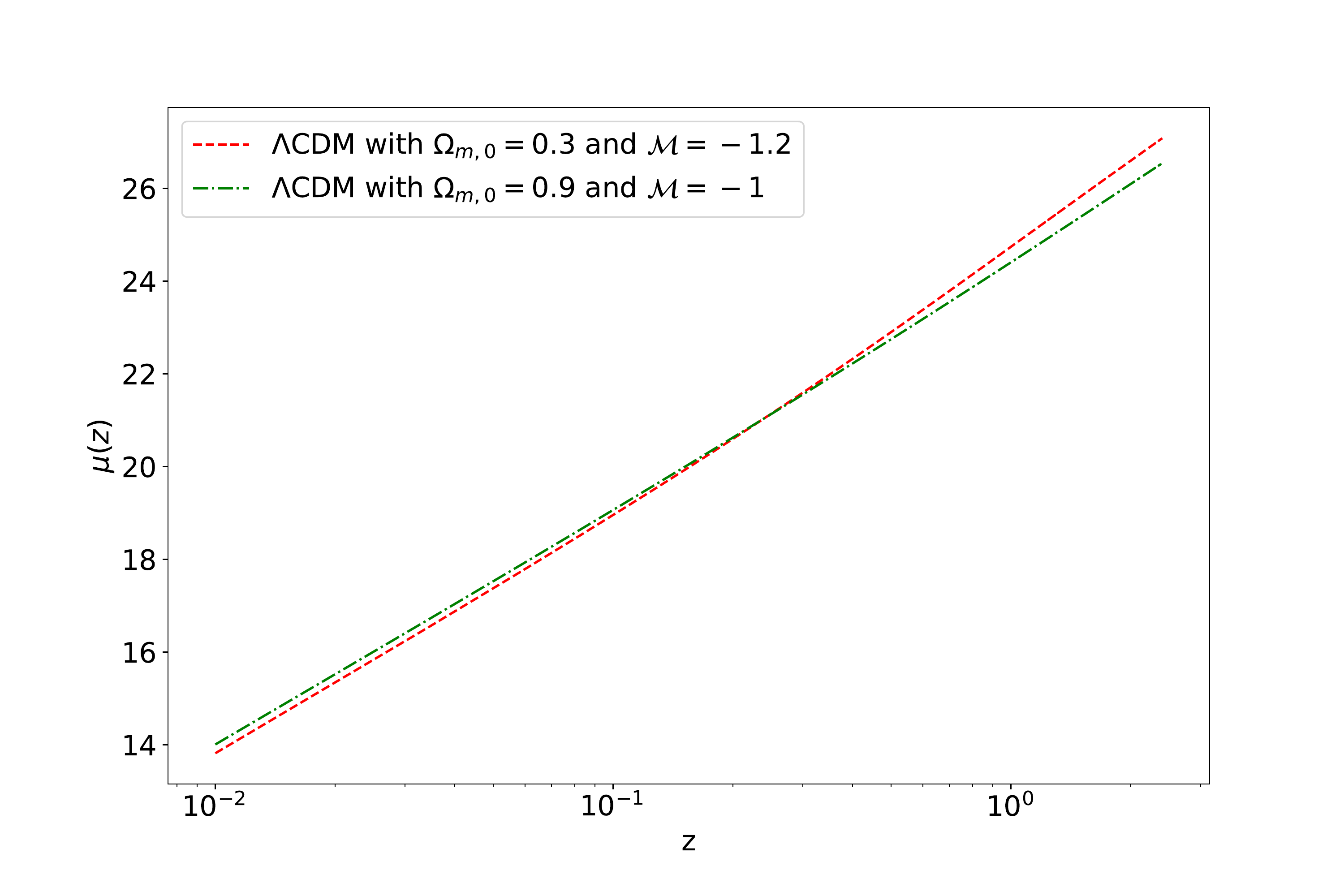}\\
\includegraphics[width=0.45\textwidth]{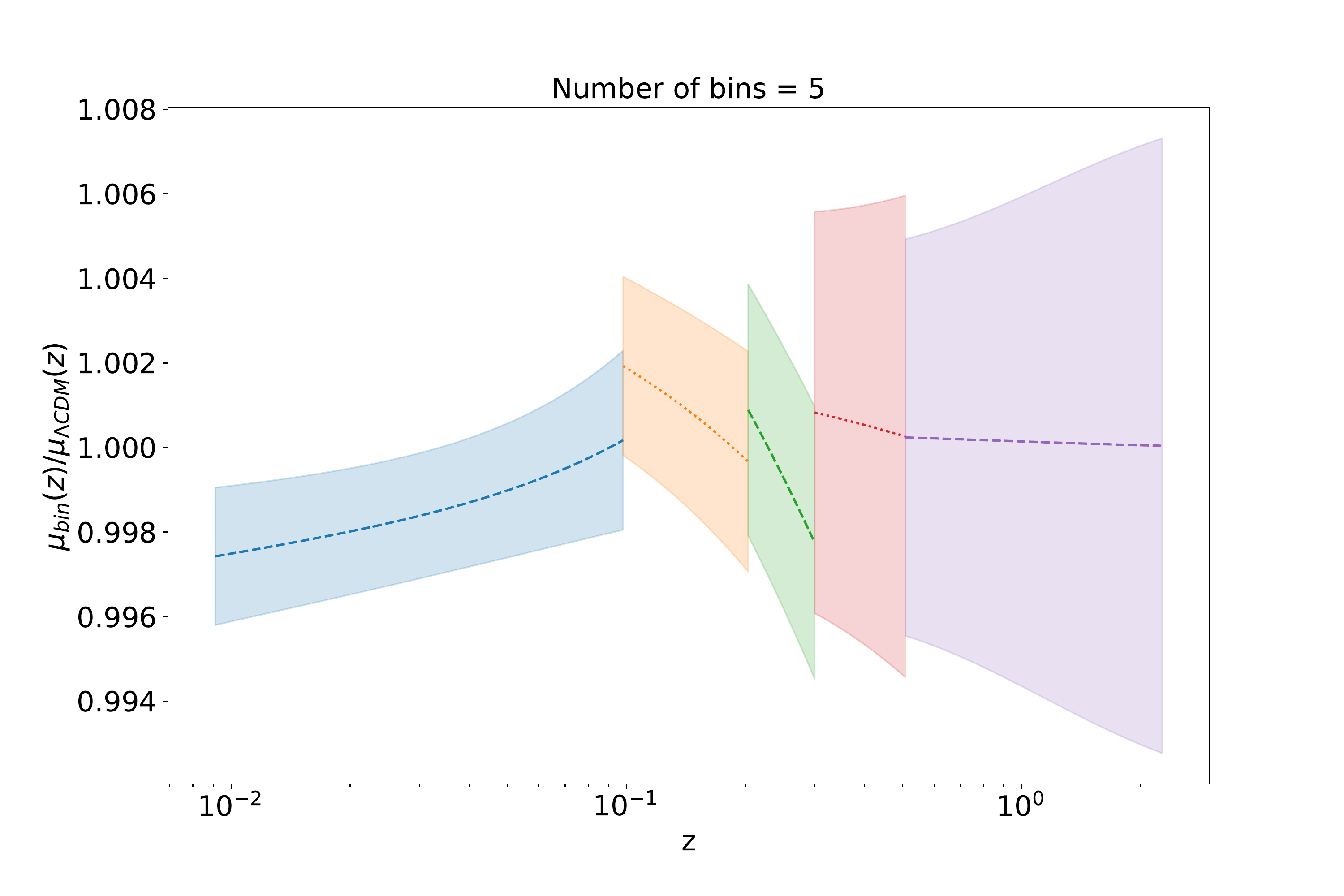}
\caption{In the top panel we show the distance modulus for two different set of values of $\Omega_{\rm m,0}-\MM$. In the bottom panel we show the reduced distance modulus for the second binning scheme normalised with the best fit of the $\Lambda$CDM model}
\label{fig:binned-mus}
\end{figure}

For the second binning scheme, we note that the matter density $\Omega_{\rm m, 0}$ experiences large fluctuations when we increase the number of bins. Firstly, the value of $\Omega_{\rm m, 0}$ decreases as the first bin decreases in width, going from $0.32$ if the bin width is  $\approx 0.25$, to $0.04$ when the bin width is  $\approx 0.1$. However, considering the 5 bins case, the matter density increases at $z\approx 0.25$ assuming values close to unity, whereas it decreases for large redshifts. 
The increasing value of the matter density is compensated by the reducing value of the absolute magnitude. 
This peculiar effect is rather an artefact than a feature. The distance modulus is a function of the luminosity distance: at the very low redshifts $d_{L}(z)$ is almost insensitive to the value of the $\Omega_{\rm m, 0}$ due to the integral nature of the luminosity distance; whereas, at high redshifts the luminosity distance decreases when the matter density increases as the Universe does not expand fast enough. However, in order to match the data, the distance modulus can adjust itself by increasing the value of $\MM$. This is exactly what happens. A further analysis shows that we can adjust both $\Omega_{\rm m,0}$ and $\MM$ to match the data. 

In the top panel of Fig.~\ref{fig:binned-mus} we show the distance modulus for the two different set of parameters $\Omega_{\rm m,0}-\MM$: the red dashed curve has been evaluated assuming a $\Omega_{\rm m,0}=0.3$ and $\MM = -1.2$ in order to match the best fit found for $\Lambda$CDM, whereas the green dot-dashed line has $\Omega_{\rm m,0}=0.9$ and the $\MM=-1$ to mimic the best fit found in the second binning scheme in the third bin. The two curves intersect around $z\approx 0.25$ which is exactly the center of the third bin (in the 5 bins case). This {\em degeneracy} is the responsible of the best fit found. In the lower panel of Fig.~\ref{fig:binned-mus} we show the reduced distance modulus for the second binning scheme with 5 bins normalised to the $\Lambda$CDM. Even though the best fit values found might point towards a new appealing physics, the distance modulus is still consistent with $\Lambda$CDM well within the $1\sigma$ errors. The errors have been evaluated using a simple propagation errors technique.

\section{Conclusions \label{sec:conclusions}}

In this paper we consider different methods to account for a varying absolute magnitude of SNIa over redshifts. We find that all the models, i.e. phenomenological functions of $\MM(z)$, modified gravity models and binning of the data, are still consistent with the $\Lambda$CDM model. 

In particular, by parameterizing $\MM(z)$ as a Taylor expansion and adding two extra parameters $\MM_0$ and $\MM_1$, we obtain results that are very close to the expected values in the $\Lambda$CDM scenario. For model 1 to 4, we perform the $\chi^2$ analysis allowing both $\MM_0$ and $\MM_1$ to vary and marginalizing over them.

For the modified gravity models, i.e. model 5 and model 6 we detect no deviations from the expected value of $g_{\rm a}=0$ and GR, by considering either the SnIa data on the their own or in combination with CMB and BAO observations. On a statistical level, we find however, that both models are on an equal footing with the \lcdm model. On the contrary, in the case of the non-homogeneous LTB model, i.e. model 7, we find that it performs significantly worse than the \lcdm model with a $\delta \chi^2 \in [7,50]$, thus not being favored by the data.

We also bin the data using two different schemes and four different binning cases. In the first binning scheme we bin only the absolute magnitude $\MM_i$ and we do not find any deviation from the constant $\MM$; whereas, for the second binning scheme we assumed also the matter density to vary with redshift in order to increase the level of complexity.  We find that there are large fluctuations in the best fit values of the parameters. However, we also realise that, with the actual data, it is possible to adjust the parameters of the models in order to mimic perfectly the distance modulus. This is reflected clearly in the top panel of Fig.~\ref{fig:binned-mus} where we assume two different values of $\Omega_{\rm m,0}$ and $\MM_i$ and we show that their corresponding distance moduli intersect at $z\approx 0.25$. This {\em degeneracy} is the responsible for the large fluctuations found point a degree of freedom on the set $\Omega_{\rm m,0}-\MM_i$. Finally, a similar approach of binning the data can be found in \cite{Nesseris:2014vra}. 

To summarize, we tested whether the absolute magnitude of the Pantheon SNIa data could be redshift dependent due to new physics, by considering several different approaches. Specifically, we considered a plethora of models ranging from Taylor expansions of the absolute magnitude, to modified gravity and cosmic void (LTB) models, but also binning methods. We found that all cases are consistent with the absolute magnitude being constant, thus ruling out the possibility it is affected by new physics.

\section*{Acknowledgements}
The authors would like to thank P.~Fleury for useful discussions. SN acknowledges support from the research project  PGC2018-094773-B-C32 and the Centro de Excelencia Severo Ochoa Program SEV-2016-0597 and the Ram\'{o}n y Cajal program through Grant No. RYC-2014-15843.
DS acknowledges financial support from Fondecyt Regular project number 1200171. CB acknowledges financial support from Swiss National Science foundation at the early stage of this work.\\

\appendix

\section{Detailed derivation of the SnIa $\chi^2$}\label{sec:appendix1}

In this section we report the details of the calculations of the $\chi^2$ used in the analysis, when we have an evolving absolute magnitude. Our new results are extensions of the marginalization calculations of \cite{Nesseris:2005ur,Nesseris:2004wj,Lazkoz:2005sp,Nesseris:2006ey,Sanchez:2009ka} and those of Appendix C of Ref.~\cite{Conley:2011ku}. Then, the full $\chi^2$ is given by 
\be
\chi^2 = \left(\DD^i - \MM(z^i)\right) C^{-1}_{ij}\left(\DD^j - \MM(z^j)\right)\,,%\nonumber
\ee
where $C_{ij}$ is the covariance matrix of the data. 
The above expression can be expanded in terms of the new data vector containing both $\MM_0$ and $\MM_1$
\bea
&\left[\DD^i - \MM_0 I^i -\MM_1f(z^i)\right]^T C^{-1}_{ij}\left[\DD^j -  \MM_0I^j-\MM_1f(z^j)\right]&  \nonumber \\
&= \DD^iC^{-1}_{ij}\DD^j-\MM_0\DD^i C^{-1}_{ij}I^j-\MM_1\DD^iC^{-1}_{ij}f(z^j) &\nonumber \\
&-\MM_0 I^i C^{-1}_{ij}\DD^j + \MM_0^2I^i C^{-1}_{ij}I^j + \MM_0\MM_1I^i C^{-1}_{ij}f(z^j)&\nonumber \\ 
&-\MM_1 f(z^i) C^{-1}_{ij}\DD^j+ \MM_0\MM_1 f(z^i) C^{-1}_{ij}I^j &\nonumber \\
&+\MM_1^2 f(z^i) C^{-1}_{ij}f(z^j)&
\label{eq:chi2-firstcalc}
\eea
where $I^i$ is the unit vector, $f(z^i)$ is a function vector evaluated at each redshift of the data sample, $\DD^i$ is the data vector.  
Eq.~\eqref{eq:chi2-firstcalc} can be written in a more compact form as:
\bea
\chi^2 &=& A\MM_0^2-2\MM_0 B+D\MM_1^2\nonumber \\
&-&2F\MM_1+2\MM_0\MM_1H+E 
\label{eq:chi2expr}
\eea
where the terms in the $\chi^2$ are given by
\bea
A &=& I^i C^{-1}_{ij}I^j\nonumber\\
B &=& \DD^i C^{-1}_{ij}I^j \nonumber\\
D &=& f(z^i) C^{-1}_{ij}f(z^j)\nonumber\\
E &=& \DD^i C^{-1}_{ij}\DD^j \nonumber\\
F &=& \DD^i C^{-1}_{ij}f(z^j) \nonumber\\
H &=& I^i C^{-1}_{ij}f(z^j)\,.\nonumber
\eea
Now, we are in the position to integrate over $\MM_0$ and $\MM_1$, and we find 
\bea
\chi^2_{\rm marg,\,\MM} &=& -2\ln\left[\int_{-\infty}^{\infty}{\rm d}\MM_1\int_{-\infty}^{\infty}{\rm d}\MM_0 \exp[-\chi^2/2]\right] \nonumber \\
&=& \frac{A F^2+B^2 D-2 B F H}{H^2-A D} \nonumber \\
&+& \ln \left(A D-H^2\right)+E-2 \log (2 \pi )\,.
\label{eq:app-marg-chi2}
\eea
Alternatively, we can also minimize the $\chi^2$ given by Eq.(\ref{eq:chi2expr}) in terms of $(M_0,M_1)$. The best-fit parameters are 
\bea
\MM_0 &=& \frac{BD-FH}{AD-H^2}  \\
\MM_1 &=& \frac{AF-BH}{AD-H^2}
\eea
and the minimized $\chi^2$ is
\be
\chi^2_{\rm min}=\frac{A F^2+B^2 D-2 B F H}{H^2-A D}+E\,.
\label{eq:app-minim-chi2}
\ee
We notice that $\chi^2_{\rm marg}=\chi^2_{\rm min}+\ln \left(A D-H^2\right)-2 \log (2 \pi )$.

Note that while searching for the minimum for model 3 and model 4, $\alpha$ might assume a value very close to zero, and as a consequence we have $A=D=F=H$ and the $\chi^2_{\rm marg}$ is singular. We can solve this problem by assuming that if the value of $|\alpha|<0.001$ then the marginalized $\chi^2$ reduces to the case of $\MM = {\rm constant}$, i.e. 
\be
\chi^2 = E + \ln \frac{A}{2\pi} -\frac{B^2}{A}\,.\nonumber
\ee

\section{Binning the data \label{sec:appendix2}}

\begin{figure}[!t]
\centering
\vspace{.1cm}\includegraphics[width=0.52\textwidth]{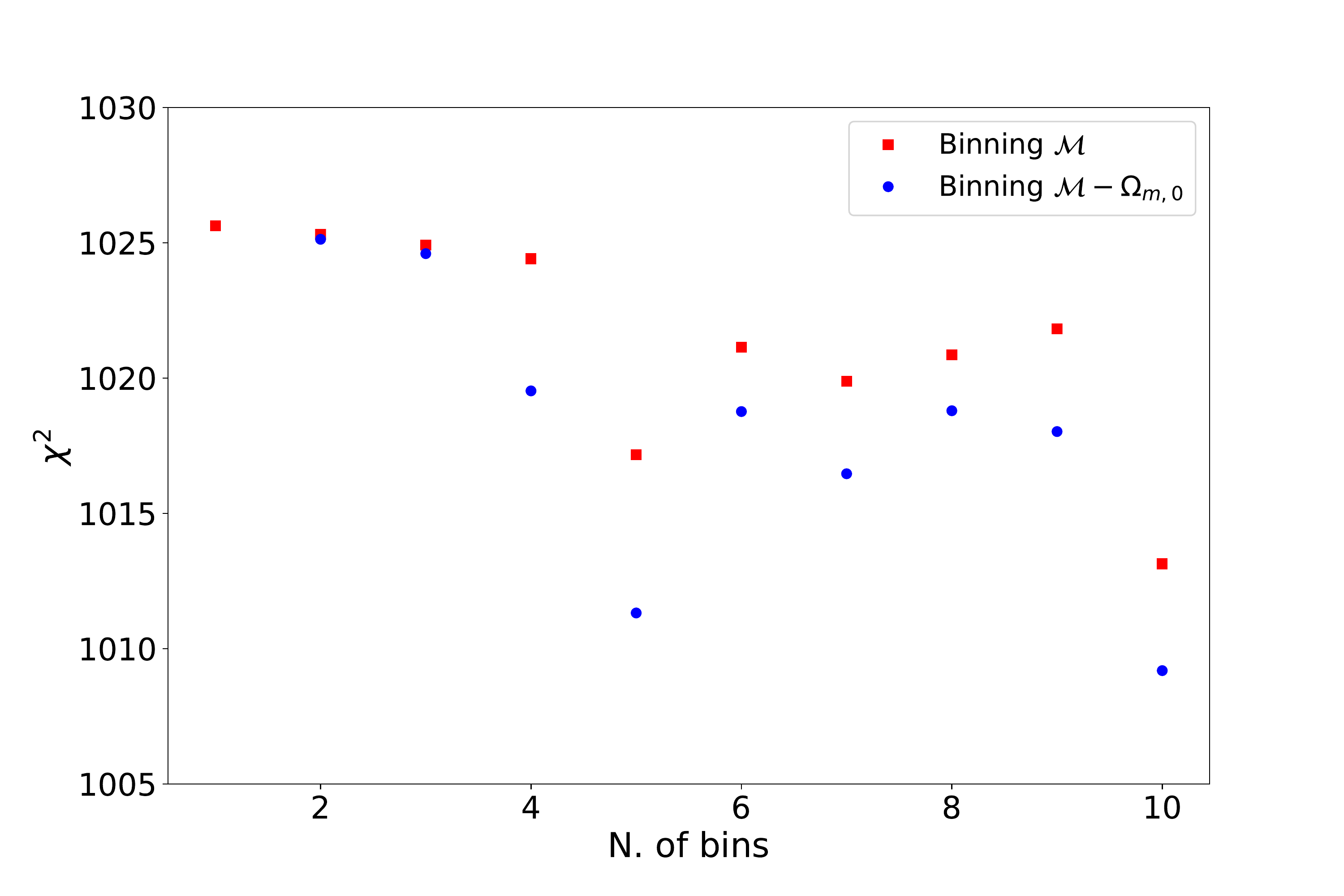}
\caption{We show the $\chi^2$ for the two different binning schemes and different number of bins. We assume a flat $\Lambda$CDM model. }
\label{fig:binned-mus-10}
\end{figure}

As an interesting test we also increased the number of bins to understand the behaviour of the $\chi^2$ found. In Tables~\ref{tab:bf-flat-binned} and \ref{tab:bf-flat-binned_mu_om} the $\chi^2$ decreases when the number of bins increases. Although this behaviour can be easily explained by the fact that we are increasing the number of parameters and hence we are overfitting the data. In Fig.~\ref{fig:binned-mus-10} we show only the values of the $\chi^2$ for the two binning schemes considered in this work. As it can be seen, if the number of bins increases the $\chi^2$ start to fluctuate and it does not decrease monotonically. Something similar happens if one considers a polynomial with increasing number of parameters due to overfitting. Then, the $\chi^2$ exhibits a very similar behavior, see Fig. 1 (top left) in  Ref.~\cite{Nesseris:2012cq}.

\bibliography{gSNIa_mu_z}

\end{document}